\documentclass[%
 aip,
 amsmath,amssymb,
 reprint,%
]{revtex4-2}
\usepackage{lipsum}
\usepackage{graphicx}% Include figure files
\usepackage{dcolumn}% Align table columns on decimal point
\usepackage{bm}% bold math
\usepackage{amsmath}
\usepackage{upgreek}
\usepackage{xcolor}
\usepackage{hyperref}% add hypertext capabilities

\begin{document}

\title[The following article has been submitted to Applied Physics Letters (APL). It will be found at APL after publication.]{Platinum contacts for 9-atom-wide armchair graphene nanoribbons}
\author{Chunwei Hsu}
\affiliation {\small \textit Kavli Institute of Nanoscience, Delft University of Technology, Lorentzweg 1, Delft 2628 CJ, The Netherlands}
\author{Michael Rohde}
\affiliation {\small \textit Kavli Institute of Nanoscience, Delft University of Technology, Lorentzweg 1, Delft 2628 CJ, The Netherlands}
\author{Gabriela Borin Barin}
\affiliation{Nanotech@surfaces Laboratory, Empa, Swiss Federal Laboratories for Materials Science and Technology, 8600 Dübendorf, Switzerland}
\author{Guido Gandus}
\affiliation{Nanotech@surfaces Laboratory, Empa, Swiss Federal Laboratories for Materials Science and Technology, 8600 Dübendorf, Switzerland}
\affiliation{Integrated Systems Laboratory, Department of Information Technology and Electrical Engineering, ETH Zurich, CH-8092 Zurich, Switzerland}
\author{Daniele Passerone}
\affiliation{Nanotech@surfaces Laboratory, Empa, Swiss Federal Laboratories for Materials Science and Technology, 8600 Dübendorf, Switzerland}
\author{Mathieu Luisier}
\affiliation{Integrated Systems Laboratory, Department of Information Technology and Electrical Engineering, ETH Zurich, CH-8092 Zurich, Switzerland}
\author{Pascal Ruffieux}
\affiliation{Nanotech@surfaces Laboratory, Empa, Swiss Federal Laboratories for Materials Science and Technology, 8600 Dübendorf, Switzerland}
\author{Roman Fasel}
\affiliation{Nanotech@surfaces Laboratory, Empa, Swiss Federal Laboratories for Materials Science and Technology, 8600 Dübendorf, Switzerland}
\affiliation{Department of Chemistry and Biochemistry, University of Bern, 3012 Bern, Switzerland}
\author{Herre S. J. van der Zant}
\affiliation {\small \textit Kavli Institute of Nanoscience, Delft University of Technology, Lorentzweg 1, Delft 2628 CJ, The Netherlands}

\author{Maria El Abbassi}

\affiliation {\small \textit Kavli Institute of Nanoscience, Delft University of Technology, Lorentzweg 1, Delft 2628 CJ, The Netherlands}

\date{\today}

\begin{abstract}
        Creating a good contact between electrodes and graphene nanoribbons (GNRs) has been a longstanding challenge in searching for the next GNR-based nanoelectronics. This quest requires the controlled fabrication of sub-20 nm metallic gaps, %a clean transfer of the GNRs  (onto electrodes without the contamination from organic polymers and damages on the GNRs in the process of device fabrication)
        a clean GNR transfer minimizing damage and organic contamination during the device fabrication, as well as work function matching to minimize the contact resistance. Here, we transfer 9-atom-wide armchair-edged GNRs (9-AGNRs) grown on Au(111)/mica substrates to pre-patterned platinum electrodes, yielding polymer-free 9-AGNR field-effect transistor devices. Our devices have a resistance in the range of $10^6$ to $10^8$ $\Omega$ in the low-bias regime, which is 2 to 4 orders of magnitude lower than previous reports. Density functional theory (DFT) calculations combined with the non-equilibrium Green's function method (NEGF) explain the observed p-type electrical characteristics and further demonstrate that platinum gives strong coupling and higher transmission in comparison to other materials such as graphene.
\end{abstract} 
\maketitle
%\section*{Introduction}

Atomically precise (GNRs) are a family of graphene-based quantum materials which have been predicted to host exotic physical properties and potential electronic applications~\cite{Saraswat2021}. Depending on their sizes and terminations, they can manifest magnetically ordered edges~\cite{Munoz-Rojas2009,Magda2014,Li2016}, tunable band-gaps~\cite{Son2006,Schwierz2010,Sevincli2008} or high-charge mobility~\cite{Baringhaus2014}. Properties such as bandgap tunability, topological properities as well as edge magnetism ~\cite{Nakada1996,Cao2017,Groning2018} and others are intrinsic to GNRs and only appear when atomic precision in the synthesis is achieved. To translate tehse properties into devices, transfer of the ribbons to an appropriate substrate and create a good electrical contact between the GNRs and the electrodes.

Two main contact approaches have been investigated so far in the literature. One is achieved by the direct deposition of electrodes on top of GNRs with lithographic tools~\cite{Bennett2013,Chen2016,Llinas2017,Fairbrother2017,Mutlu2021}. GNR devices fabricated with this top-contact approach show high, non-Ohmic contacts, and in some cases, the current is limited by the contact resistance~ \cite{Bennett2013,Chen2016,Fairbrother2017}. This indicates a poor contact of the GNRs with the possible presence of a large Schottky barrier. %From a technological point of view, this limits the performance and reliability of the GNR devices; fundamentally, it poses a question if the intrinsic properties of GNRs are correctly measured.
Additionally, these top-contact GNR devices can suffer from resist contamination and heating during metal evaporation in the lithography process. This is particularly destructive for the GNRs with reactive edges, such as spin-polarized edges and topologically protected edgestates~\cite{Ruffieux2016,Groning2018,Keerthi2017}. 

Another approach for contacting GNRs is by transferring GNRs onto pre-patterned electrodes. %This approach prevents polymer contamination and damages during lithography. %, which is of particular importance for the device integration of GNRs with reactive edges~\cite{Ruffieux2016,Groning2018,Keerthi2017}.
For GNRs grown on Au-mica substrates, a polymer-free transfer of 9-AGNRS has been optimized and used in previous reports of Refs.~\citenum{Llinas2017,BorinBarin2022} and Ref.~\citenum{Braun2021} with Pd and graphene electrodes, respectively. However, in the case of Pd nanogaps, large Schottky barriers limited the transport through these devices. Likewise, graphene electrodes did not solve the issue of contact resistance and introduced more uncertainties related to the fabrication, i.e., not well-defined gap sizes and lithography-related PMMA residues on the electrodes, the latter being a known concern for graphene devices ~\cite{Li2018}.

%The first report shows low resistance of a few M$\Omega$ with a large channel length of hundreds of nanometers. The latter report can suffer from the polymer residue during the graphene etching, as it is known that PMMA on graphene surface is hard to be removed~\cite{Li2018}. With a thin PMMA resist layer and the cold development method, we can achieve Pt nanogap as small as 20~nm with an aspect ratio of more than 100. This allows the possibility to measure end-to-end connected GNRs as they are in the length of few tens of nanometers; moreover, the large aspect ratio provides a high device yield as many 9-AGNRs can be connected in the same junction. Similar nanogaps with large aspect ratio can also be achieved with other methods but requires more elaborated techniques such as chromium oxide mask~\cite{Fursina2008}. It thus grants us a straight forward way to create nanogap with metal electrodes with the size of the 9-AGNR's lengths, in the orders of few tens of nanometers.
\begin{figure}[b]
    \centering
    \includegraphics{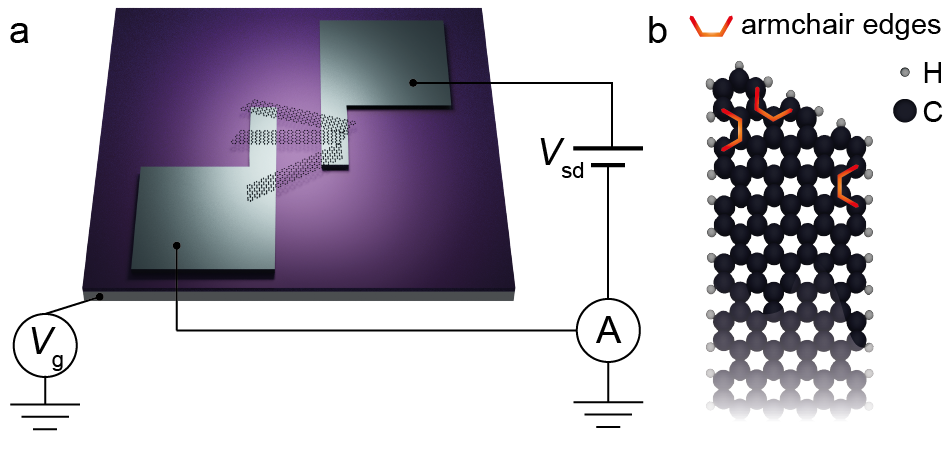}
    \caption{Device schematics: a, 9-AGNR field effect transistor device with Pt electrodes and $\rm SiO_2$ as the back gate oxide. b, Atomic structure of 9-AGNR. The armchair termination is indicated by the red lines in the structure.}
    \label{fig:Schematic}
\end{figure}

\begin{figure*}[!t]
    \centering
    \includegraphics{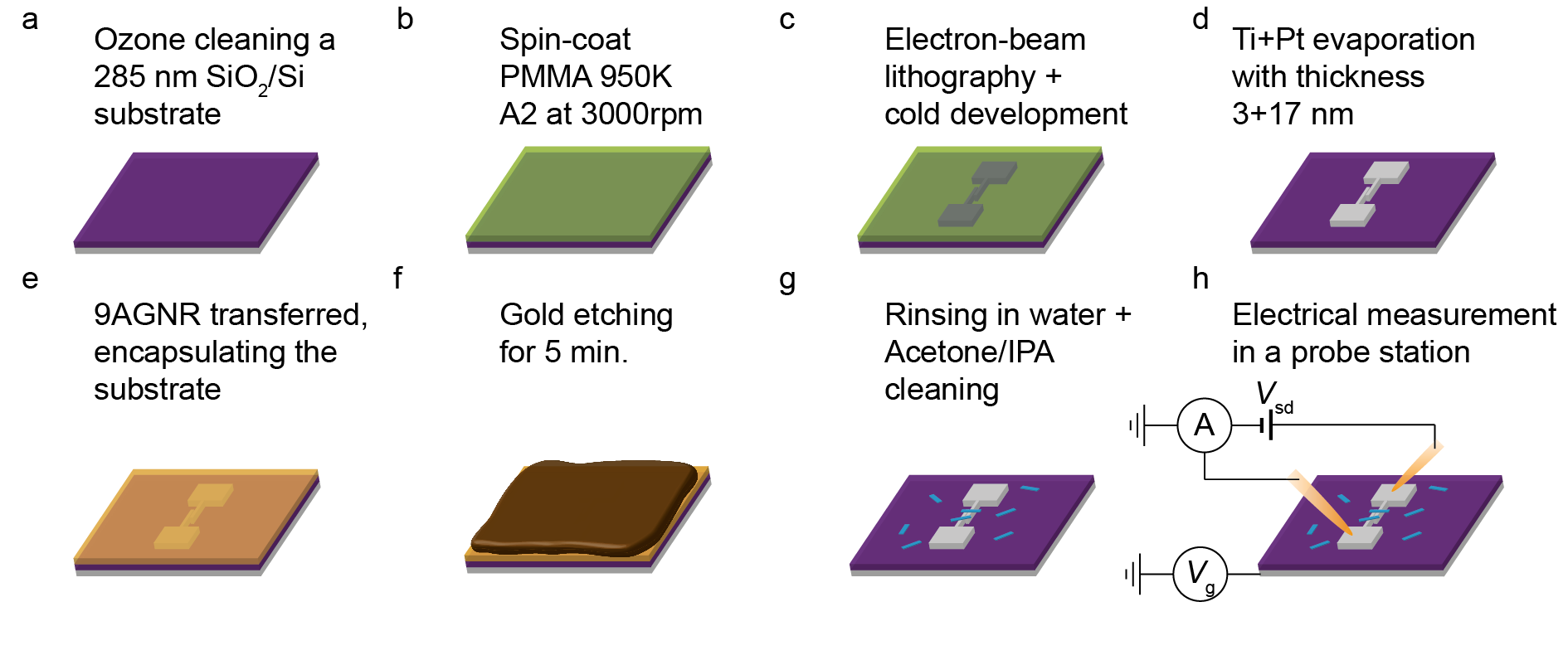}
    \caption{Fabrication steps: a-d, The pre-patterning of the Pt nanogap. e-g, Polymer-free transfer of 9-AGNR. h, Electrical measurements in a probestation. The blue lines represent the 9-AGNRs.}
    \label{fig:Fabrication}
\end{figure*}

In this letter, we study 9-AGNRs junctions with pre-patterned Pt electrodes forming nanogap ranging from 20 to 100~nm in width and 1~$\upmu$m in length. We transfer the 9-AGNRs after the nanogap fabrication and show GNR devices with low-bias (200 mV) Ohmic resistance in the range of $10^6$ to $10^8$ $\Omega$, orders of magnitude lower than the previous reports~\cite{Bennett2013,Chen2016,Llinas2017,Fairbrother2017,Braun2021}. Approximately $100\%$ device yield and low resistance are realized as a result of a cleaner device fabrication process compared to the previous top-contact approach, where GNRs are subjected to polymer contamination and high process temperatures. 
With the field effect transistor geometry, we further demonstrate the high transmission in the Pt-GNR-Pt junctions with p-type transport properties, concluded from gate-dependent measurements. 
These observations are rationalized by density functional theory and non-equilibrium Green's function formalism  (DFT+NEGF) calculations, estabilishing platinum as an excellent material for contacting 9-AGNRs. 
%Yet, there has been a technological difficulty in electronically contacting these small entities with a top-contact approach, due to their small sizes (few to tens of nanometers) and possible damage on the GNRs during the fabrication processes. 

%Recently, several groups have experimented in contacting the GNRs by depositing metallic electrodes directly~\cite{Bennett2013,Chen2016,Llinas2017,Fairbrother2017,Mutlu2021}. However, these top-contact junctions show highly non-ohmic contacts, and mostly exhibit small currents in the range of a few nA after the application of a source-drain voltage from a few to tens of volts. This highly resistive non-ohmic contact indicates a poor metalization of the GNRs with the possible presence of a large Schottky barrier. From a technological point of view, this limits the performance and reliability of the GNR devices; fundamentally, it poses a question if the intrinsic properties of GNRs are correctly measured. As pointed out by more recent research, such device performance can suffer from organic contamination on the GNR films during the lithography process, leading to the non-Ohmic, highly resistive bahaviors~\cite{Richter2020,Braun2021}. In addition, GNRs with novel properties, such as spin polarized edges and topologically protected edges states~\cite{Ruffieux2016,Groning2018} are much more reactive than the simpler armchair GNRs and the top-contact fabrication techniques can introduce damage causing the loss of their electronic properties.

%\section*{Experimental and Fabrication}
We employ a field-effect transistor geometry to electrically characterize 9-AGNRs in a vacuum probe station. A schematic device lay-out is illustrated in Fig.~\ref{fig:Schematic}a. With this geometry we measure the current-voltage (\textit{IV}) characteristics of 9-AGNRs as well as their gate dependence ($IV_\mathrm{g}$). The GNRs are transferred onto pre-patterned Pt gaps on a SiO$\rm _2$/Si substrate, where the Si wafer is used as a global back-gate electrode. The 9-AGNRs form transport channels by bridging the pre-patterned lithographically defined Pt nanogaps. The atomic structure of 9-AGNRs is also shown in Fig.~\ref{fig:Schematic}b, where the four sides of the GNRs are armchair-terminated.  

To form a clean 9-AGNR-electrode interface, we pattern Pt electrodes prior to introducing the GNRs, thus avoiding organic contamination of the junction. The device fabrication steps are shown in Fig.~\ref{fig:Fabrication}. We use a SiO$_2$/Si substrate with a thermal oxide thickness of 285~nm. %The choice of 285~nm is to have a thick enough oxide layer which can be used for contacting the device to avoid a leak to the gate, while keeping a sufficient gate coupling to the GNR. 
The substrates are first cleaned with acetone and isopropyl alcohol (IPA) for 5 minutes each to remove organic residues on the surface. Subsequently, the substrate is cleaned with an oxygen plasma at a power of 300~W for 3 minutes. After the cleaning, the substrate is spin-coated with PMMA 950K A2 (MicroChem) at 3000 rpm and baked at 180~$^\circ $C for 3 min on a hot plate. This gives a resist thickness of about 80~nm.

The nanogaps with various widths (20-100 nm by design) are patterned by EBPG5000+ (Raith) with an acceleration voltage of 100~kV. To form well-defined nanogaps, a high dose at 2100 $\upmu \rm C/cm^2$ is chosen together with a cold development technique. The nanogap structure is developed in IPA:MIBK (3:1) at a temperature of -20$^{\circ}$C for 3 min. Afterwards, the electrodes are made by electron-beam evaporation of 3~nm of Ti at a rate of 0.5~\AA/s and 17~nm of Pt at a rate of 1~nm/s, followed by a lift-off process in hot acetone at 50~$^{\circ}$C.  With this thin PMMA resist layer and the cold development method, we achieve Pt nanogap as small as 20~nm with an aspect ratio of more than 100. This allows the possibility to measure end-to-end connected GNRs as they are in the length of few tens of nanometers (see Fig.~S1a for a STM image of 9-AGNRs on Au(111)). Moreover, the large aspect ratio provides a high device yield as several 9-AGNRs can be connected in the same junction. Similar nanogaps with large aspect ratio can also be achieved with other methods but require a more elaborated technique such as a chromium oxide mask~\cite{Fursina2008}.

To transfer the 9-AGNRs onto the pre-patterned substrate, we follow the polymer-free transfer process described elsewhere~\cite{Cai2014,Fairbrother2017,BorinBarin2019}. In short, an Au film containing 9-AGNRs delaminates itself from its mica substrate when placed onto an aqueous HCl solution (Fig.~\ref{fig:Fabrication}e). Afterwards, a pre-patterned substrate is used to pull out the free-standing Au film from the diluted HCl solution. To remove the Au film from the 9-AGNRs, the substrate with Au film is covered with a gold etchant for 5 min., as shown in Fig.~\ref{fig:Fabrication}f, and subsequently rinsed with deionized water and cleaned with acetone and IPA. This transfer process preserves the 9-AGNR quality as no peak shift in the Raman spectra of 9-AGNRs was observed, before and after the transfer (Fig.~S1b). Afterwards, the sample is mounted in a vacuum probe station at room temperature for electrical characterization.
\begin{figure}[t]
    \centering
    \includegraphics[width=\columnwidth]{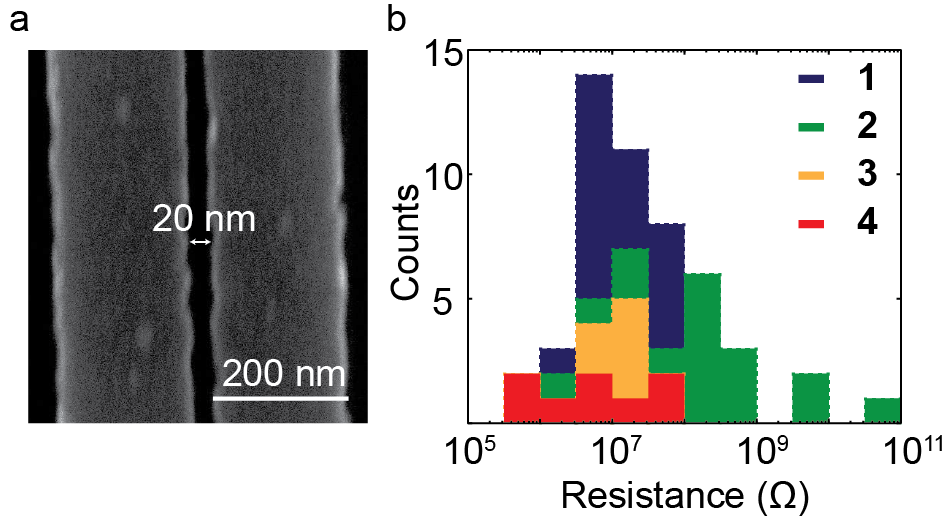}
    \caption{a, Scanning electron microscopy image of Pt nanogap, with a feature size of 20~nm. b, Resistance determined at a bias voltage of 50~mV for four different samples with several devices on each of them.}
    \label{fig:Characterization}
\end{figure}
 Figure~\ref{fig:Characterization}a shows a scanning electron microscopy (SEM) image of a typical Pt nanogap with a feature size of 20~nm. 9-AGNRs with an average length of 45~nm and maximum lengths up to 100~nm are transferred onto these nanogaps, forming Pt-9-AGNR-Pt junctions. Here, we present electrical measurements of four different substrates, each contains multiple devices. Figure~\ref{fig:Characterization}b summarizes the resistance of these junctions at 50~mV with all nanogap sizes.The junctions have a most probable low-bias resistance around $10^7~\Omega$, orders of magnitude lower than that in previous reports~\cite{Bennett2013,Chen2016,Llinas2017,Fairbrother2017,Braun2021}.  We also compare the low-bias resistance between junctions, made with a polymer-free transfer and PMMA-assisted transfer techniques on the same substrate (see Fig.~S2). Consistently, the PMMA-assisted transferred junctions show a resistance that is 1 to 2 orders of magnitude higher, showing a clear influence of the use of PMMA on the transport properties, possibly due to different doping levels and contamination at the interfaces.

It is worth noting that the yield is nearly 100\% for electrically conducting junctions, i.e., $R<10~\rm G\Omega$ after GNR transfer.  All devices were previously verified to be insulating. The large 9-AGNR/Pt contact area (1~$\upmu$m in vertical direction of Fig.~\ref{fig:Characterization}a) may be of importance in this observation. A high yield benefiting from a large junction contact area was also observed previously, even in devices with $\upmu $m-size gaps~\cite{Richter2020}. This implies that the Pt-9-AGNR junctions comprise a network of GNRs, where a distance-dependent resistance is expected~\cite{Richter2020}. However, with the large spread in the resistance distribution, we do not observe significant difference in the low-bias resistance for devices with different nanogap sizes ranging from 20~nm to 100~nm (see Fig.~S3).

\begin{figure}[!t]
    \centering
    \includegraphics{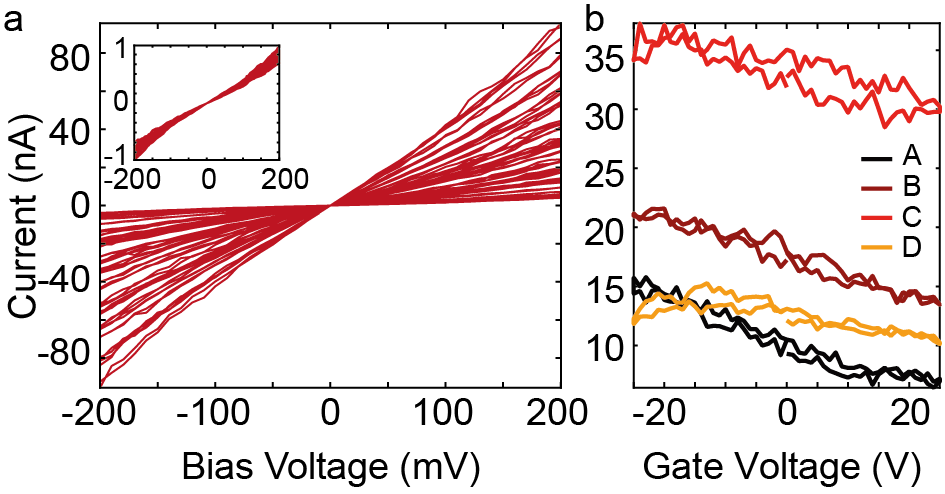}
    \caption{a, \textit{IV} characteristics of 36 junctions in sample 1 within a bias voltage range of $\pm$ 200~mV. The inset shows the normalized IV characteristics. b, Gate traces of 4 different junctions in sample 1, taken at a bias voltage of 100~mV.}
    \label{fig:IVs_Gate}
\end{figure}

To gain more insight into charge transport in the 9-AGNR devices, we show the \textit{IV} characteristics of sample~1 in Fig.~\ref{fig:IVs_Gate} (see Fig.~S4 for other samples). The current shows a linear dependence on the bias voltage within the range of $\pm$200~mV. To probe the linearity of the \textit{IV} characteristics, we have applied a bias voltage up to 1~V, shown in Fig.~S5 for sample~2. This bias limit is chosen to prevent the possible creation of filamentary paths in silicon oxide, which can occur at a few Volts 
 applied across thin oxide layers~\cite{Mehonic2018}. In this case, we observe small nonlinearity taking place typically around a few hundred mV. 
The low resistance and nearly Ohmic $IV$s of these pre-patterned Pt-9-AGNR-Pt junctions can be a result of a better work function matching in comparison to previously investigated electrode materials~\cite{Bennett2013,Chen2016,Llinas2017,Fairbrother2017,Braun2021}.

Figure~\ref{fig:IVs_Gate}b shows the gate dependent current at a bias voltage of 100~mV. From the ratio of the maximum and minimum current for gate voltages ranging from -20~V to 20~V, the on-off ratio is determined: $R_\textrm{on-off}=I_{\rm max}/I_{\rm min}$. We obtain small $R_\textrm{on-off}<10$ for sample 1. The highest on-off ratio of 25 is observed in sample 3, shown in Fig.~S6d. Additional gate traces at a higher bias voltage of 1~V for sample 2 are shown in Fig.~S7, where a maximum $R_{\rm on-off}$ of 30 is observed. A crucial observation in the gate traces, consistent p-type transport, which is consistent with previous observations in 9-AGNR-based transistors~\cite{Bennett2013,Chen2016,Llinas2017,Fairbrother2017,Mutlu2021,Richter2020,Braun2021}.

\begin{figure*}[!ht]\centering
    \includegraphics{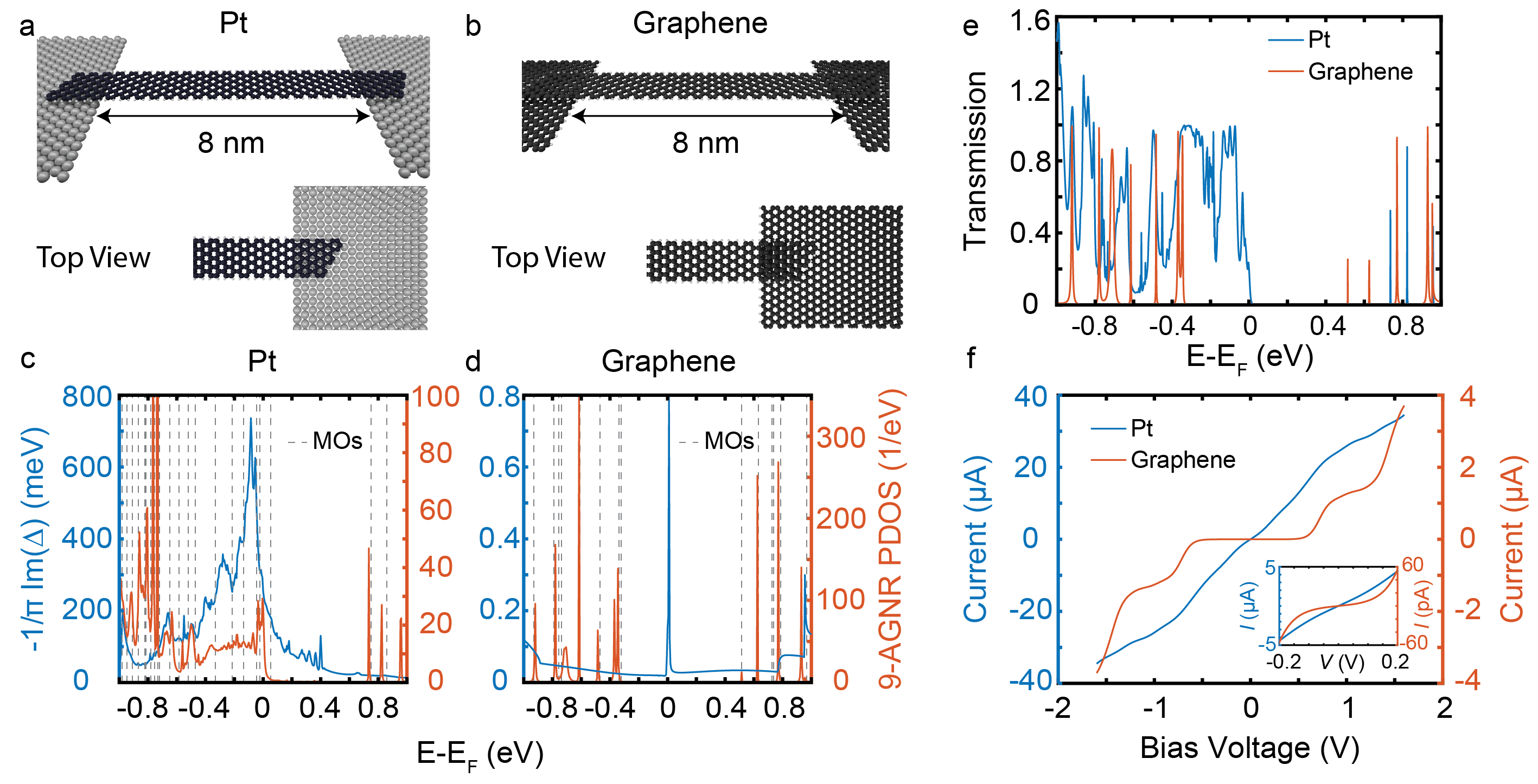}
    \caption{a-b, Atomic models for 9-AGNR nanogap junctions with platinum and graphene contacts respectively. c-d, Hybridization function between the GNR and the metallic contacts as a function of energy. The hybridization with platinum is orders of magnitude larger than with graphene, which is reflected in the "broad" PDOS below the Fermi level. For graphene, the PDOS is characterized by sharp peaks directly linkable to the MOs of the dissociated GNR indicating that the molecule is only loosely coupled to the contacts. e, Electronic transmission as a function of energy. It is characterized by narrow peaks in regions of small hybridization and approaches the idea constant value of 1 when the hybridization increases. f, Current obtained by integrating the electronic transmission in a bias window centered around the Fermi level. The current for platinum contacts shows the linear behaviour observed in experiments.}
    \label{fig:Calculation}
\end{figure*}

The small gate dependence and on-off ratio of the current illustrate the poor gate efficiency of the 9-AGNR junctions. This poor efficiency may come from three contributions: (i) The electric field is screened by the metallic electrodes between the gate and GNR, and screened between GNR and GNR. These screening effects were demonstrated by simulations previously in ref.~\citenum{Mutlu2021}, where the electrostatic potential was completely screened in a densely packed GNR film with a 
GNR separation of 1.5~nm. It was also shown that a 20~nm nanogap is screened for more than 50\% with an electrode thickness of 4~nm without the presence of GNRs~\cite{Mutlu2021}. This suggests that the densely packed 9-AGNR devices with a metal electrode height of 20~nm can be screened efficiently, leading to the observed poor gate efficiency. (ii) The low gate coupling is partially due to the low dielectric constant of the silicon oxide and the thick oxide thickness. An improvement for future experiments can be made by using a thin, high-$\kappa$ gate oxide such as $\rm HfO_2$.  (iii) The transport mechanism can be dominated by a hopping-like mechanism with an intrinsic low gate coupling. %Two important features of the hopping mechanism are the power-law distance and temperature dependence of resistance. The gap sizes used in our devices are in the same order as the GNR length, suggesting there can only be very few or single hopping sites for charge transport. Due to the large spread in the resistance distribution, the distance dependence remains inconclusive in our observations.
A temperature dependent measurement is necessary to elucidate the transport mechanism in these GNR devices, which is a subject of study for the future.

%This provides the possibility to study the intrinsic electrical properties of end-to-end connected 9-AGNRs.  As hopping-like transport has strong distance dependence, the absence of distance dependence are reasoned by two possible explanations. First, since the gap sizes are comparable to the 9-AGNR lengths, most of the charge transport happens through single ribbons or through single to few hopping sites. The other possibility is that the resistances are limited by the contact resistance, even with the clean GNR/Pt interface. Such nonlinear IVs show the presence of barriers, which can either take the form of Schottky barrier and hopping transport. The elucidation of the origin of barrier requires the temperature dependence of the IVs, which is a subject for future studies.

%\section*{Density Functional Theory}

To describe the low resistance and nearly linear IV characteristics of the 9-AGNR junctions, we employed the density functional theory + non equilibirum Green's function (DFT+NEGF) method to unveil the origin of this behavior in our 9-AGNR-Pt coupling. Details about the simulations can be found in SM.2.

We consider the atomistic models in Figs.~\ref{fig:Calculation}a and b. They are representative of a 9-AGNR contacted with platinum and graphene, respectively, with the goal of comparing both configurations. The length of the GNR is set to 11 nm and the contacts are separated by a 8-nm gap. For ribbons with atomically precise edges transport is expected to be ballistic and the conductance should not vary with the distance between the contacts~\cite{koch2012voltage}. To assess the quality of the latter we compute the hybridization strength of the GNR with the underlying electrodes and report the result in Figs.~\ref{fig:Calculation}c-d. For platinum contacts, the hybridization is large below the Fermi level, i.e., in the valence band (VB) of the GNR, but decreases above this energy. This finding is evidenced by the density of states projected onto the GNR (PDOS), which evolves from a continuum of states below in the VB to a series of discrete and narrow peaks above it, directly linkable to the molecular orbital states (MOs) of the uncoupled GNR (vertical dashed lines). 

For graphene contacts on the other hand, the hybridization remains small throughout the entire energy window relevant for transport and the PDOS strongly resembles the discrete spectrum of the dissociated GNR. This indicates that, upon contact with graphene, the electronic states in the channel remain mostly bounded. The hybridization strength is reflected in the electronic transmission (see Fig.~\ref{fig:Calculation}e), which approaches the ideal constant value of 1 of a perfectly contacted GNR in the regions of large hybridization and reduces to sharp peaks typical of resonant transport otherwise. The current calculated by applying a symmetric bias across the electrodes shows a linear dependence for platinum contacts within the range of $200$~mV, in agreement with experiments (see Fig.~\ref{fig:Calculation}f and \ref{fig:IVs_Gate}a). In our calculations, the linear dependence stems from the $p$-type character of the GNR - any bias window finds some open channels available for transport. Contrarily, the current for graphene contacts shows a non-linear, step-like behaviour typical of molecular junctions in the resonant transport regime in which the central molecule is loosely coupled to the leads. 

Our simulation results thus indicate that, upon contact with platinum, the GNR MOs strongly hybridize with the underlying material and broaden into a continuous density of channels available for transport. This in turn yields devices with low contact resistance and nearly linear IV characteristic.

% Next, we show that graphene electrodes yield devices with higher resistance due to the weaker 9-AGNR-contact couplings. We report the hybridization, transmission and current in Fig.\ref{fig:DFT-graph}. The hybridization of the $p_z$ manifold with the $\pi$ bands is orders of magnitudes smaller than with the $d$ bands of platinum. The poor coupling is reflected in the PDOS and transmission characteristics which are characterised by sharp peaks at the MOs of the dissociated GNR indicating a long life of the electrons in the channel that cannot easily escape into the contacts. As a result, the current is exponentially suppressed in the gap and gradually increases at the beginning of each GNR MO. This stepped behaviour is in fact typical of molecular junctions governed by Coulomb blocking effects in which the central molecule is loosely coupled to the leads.

% platinum has been shown to form a better electrical with GNR-like molecules.\cite{Xie2019}

%\section*{Summary}
We fabricate 9-AGNR field effect transistor devices with Pt contacts by employing a polymer-free transfer technique subsequent to the deposition of electrical contacts. The GNR devices, ranging from 20~nm to 100~nm in gap width, consistently show a low-bias resistance value, $R\approx 10^7~\Omega$, orders of magnitude lower than previous reports. Together with its nearly-Ohmic IV characteristics, the better device performance indicate that Pt electrodes with polymer-free transfer is ideal for 9-AGNR contacting. DFT+NEGF calculations demonstrate that Pt contact leads to a higher transmission than that other materials such as graphene. This not only explains the nearly linear $IV$ characteristics and $p$-type transport observed in the experiments, but also points out that Pt is a better contact material for a transparent contact interface.

\section*{Author Contributions}
\textbf{Chunwei Hsu}: Conceptualization (equal); Methodology (equal); Validation (equal); Visualization (lead); Writing – original draft (lead); Writing – review \& editing (lead).
\textbf{Michael Rohde}: Formal Analysis (lead); Validation (equal); Visualization (equal); Writing – review \& editing (equal).
\textbf{Gabriela Borin Barin}: Resources (lead); Writing – original draft (equal); Writing – review \& editing (equal).
\textbf{Guido Gandus}: Software (lead); Formal Analysis (equal); Methodology (equal); Writing – original draft (equal); Writing – review \& editing (equal).
\textbf{Daniele Passerone}: Supervision (equal); Writing – review \& editing (equal).
\textbf{Mathieu Luisier}: Supervision (equal); Writing – review \& editing (equal).
\textbf{Pascal Ruffieux}: Funding Acquisition (equal); Supervision (equal); Writing – review \& editing (equal).
\textbf{Roman Fasel}: Funding Acquisition (equal); Supervision (equal); Writing – review \& editing (equal).
\textbf{Herre~S.~J. van der Zan}: Funding Acquisition (equal); Supervision (equal); Validation (equal); Writing – review \& editing (equal).
\textbf{Maria El Abbassi}: Supervision (lead); Methodology (equal); Validation (lead); Writing – original draft (equal); Writing – review \& editing (equal).\\

\section*{Acknowledgement}
This study was supported by the EU and FET open project QuIET (number 767187). C.~H. and H.~S.~J. v.d.Z. acknowledge The Netherlands Organization for Scientific Research (Natuurkunde Vrije Programma's: 680.90.18.01). We acknowledge funding by the Swiss National Science Foundation under grant no. 200020-182015, the European Union Horizon 2020 research and innovation program under grant agreement no. 881603 (Graphene Flagship Core 3), and the Office of Naval Research BRC Program under the grant N00014-18-1-2708. We furthermore greatly appreciate the financial support from the Werner Siemens Foundation (Carbo Quant). DP, ML and GG acknowledge the NCCR MARVEL funded by the Swiss National Science Foundation (grant no. 51NF40-205602).

\section*{Data Availability Statement}

The data that support the findings of this study are available from the corresponding author upon reasonable request.

\section*{References}
%\bibliographystyle{naturemag}
%\bibliography{bib.bib}% P
%aipnum4-2.bst 2019-01-14 (MD) hand-edited version of apsrev4-1.bst

\pagebreak
\end{document}